\documentclass[aps,prd,preprint,tightenlines]{revtex4}
\usepackage{graphicx}

\begin{document}

\title{Axino Light Dark Matter and Neutrino Masses with
R-parity Violation}
\author{Eung Jin Chun}\email{ejchun@kias.re.kr}
\affiliation{Korea Institute for Advanced Study, Seoul 130-722, Korea}
\author{Hang Bae Kim}\email{hbkim@hanyang.ac.kr}
\affiliation{BK21 Division of Advanced Research \& Education in Physics,
Hanyang University, Seoul 133-791, Korea}

\begin{abstract}
Motivated by the recent observation of the 511 keV $\gamma$-ray
line emissions from the galactic bulge and an explanation for it
by the decays of light dark matter particles, we consider the
light axino whose mass can be in the $1-10$ MeV range,
particularly, in the context of gauge-mediated supersymmetry
breaking models. We discuss the production processes and
cosmological constraints for the light axino dark matter. It is
shown that the bilinear R-parity violating terms provide an
appropriate mixing between the axino and neutrinos so that the
light axino decays dominantly to $e^+ e^- \nu$. We point out that
the same bilinear R-parity violations consistently give both the
lifetime of the axino required to explain the observed 511 keV
$\gamma$-rays and the observed neutrino masses and mixing.
\end{abstract}

\pacs{14.80.Ly, 95.35.+d}
\keywords{}

\maketitle

\section{Introduction}

The Peccei-Quinn (PQ) mechanism to solve the strong CP problem
\cite{Kim:1986ax}, when combined with supersymmetry (SUSY) which is
the solution to the gauge hierarchy problem, predicts a singlet
fermion called the axino. It can be light in certain supersymmetry
breaking mechanism, and become the lightest supersymmetric particle
(LSP) providing a good candidate for the particle dark matter (DM)
in various mass ranges
\cite{Rajagopal:1990yx,GY,Chun:1999cq,Covi:1999ty,Covi:2001nw}.

Phenomenologically viable supersymmetric models are implemented with
the R-parity to assure the stability of the proton, which also
implies the stability of the LSP. However, R-parity is not dictated
from any deep theoretical principle.  The small violation of
R-parity is an attractive option for generating the neutrino masses
and mixing \cite{Hall:1983id}.  Even with the R-parity violation,
the LSP  can be cosmologically stable and it may provide an indirect
detection mechanism of the DM by leaving imprints in $\gamma$-rays from
the galactic center and in the diffuse background \cite{Kim:2001sh}.

Recent observation of 511 keV $\gamma$-rays by the SPI spectrometer
aboard the INTEGRAL satellite not only confirmed the previously
measured total flux but also revealed the morphology of the bulge
emission, which is highly symmetric and centered on the galactic
center with a full width half maximum of $\sim8^\circ$
\cite{Knodlseder:2005yq,Jean:2003ci,Knodlseder:2003sv}. The
observed emission of 511 keV $\gamma$-rays can be well explained by
$e^+e^-$ annihilations via positronium formation. But the origin
of these galactic positrons remains a mystery. Many astrophysical
sources have been suggested, including massive stars, neutron
stars, black holes, supernovae, and X-ray binaries. The generic
problem of astrophysical sources is that they have difficulty in
explaining both the total flux and the high bulge-to-disk ratio of
observed 511 keV $\gamma$-rays. Given this difficulty, suggested were
alternative explanations that light dark matter (LDM) particles
annihilating or decaying in the galactic bulge are the sources of
the galactic positrons
\cite{Boehm:2003bt,Hooper:2004qf,Picciotto:2004rp,
Ferrer:2005xv,Kawasaki:2005xj,Kasuya:2005ay,drees}.

In addition to positrons, annihilations or decays of LDM particles
produce direct $\gamma$-rays via the internal bremsstrahlung
processes. The observation of $\gamma$-rays from the galactic center
in the energy range $1-100$ MeV bounds the mass of LDM particles
to be less than about $20$ MeV \cite{Beacom:2004pe}. It was also
claimed that astrophysical sources are missing for the diffuse
$\gamma$-ray background in the energy range $1-20$ MeV from the
observed spectrum, and that direct $\gamma$-rays from annihilations
or decays of LDM particles can fit the spectrum when the produced
positrons are normalized to fit the 511 keV $\gamma$-rays from the
galactic bulge \cite{Ahn:2005ws,Ahn:2005ck}. Concerning the
annihilating LDM, its mass less than 10 MeV is practically
excluded because it leads to a much longer supernovae cooling time
which makes impossible the emission of sufficiently energetic
neutrinos observed in SN1987A \cite{Fayet:2006sa}.

In view of above observations, the axino in R-parity violating
supersymmetric models is a well-motivated candidate for the MeV
dark matter whose decay can explain the observed 511 keV line
emission from the galactic bulge as suggested by Hooper and Wang
\cite{Hooper:2004qf}.
Indeed,  R-parity violation is  required to make the axino decay
and its lifetime can be very long since its interactions are
suppressed by the PQ scale. An interesting question one may ask is
whether the same R-parity violation can also generate the observed
neutrino masses and mixing.

In this article, we show that the axino LDM scenario is consistent
with the usual mechanism of generating the neutrino masses and
mixing at tree-level through the small bilinear R-parity violating
couplings $\sim 10^{-6}$ \cite{LeeKang}.  Such small bilinear
terms turn out to induce an appropriate axino-neutrino mixing
through which the light axinos decay to  the positrons with the
right range of the lifetime \cite{Hooper:2004qf};
\begin{equation} \label{primo}
 \tau_{\rm dm} \sim {4\times10^{26} \over  m_{\rm dm}({\rm
MeV}) } \, {\rm sec} \,.
\end{equation}
This has to be contrasted to the case of \cite{Hooper:2004qf}
where the trilinear couplings $\lambda_{i11}\sim 0.1$ were
considered.

We also discuss how the MeV axino can arise, particularly, in
gauge mediated supersymmetry breaking (GMSB) schemes where the
saxion is predicted to get the mass in the range of $4-50$ GeV.
Axinos are produced thermally or non-thermally in the early
universe and the amount of axinos can be correctly adjusted for
the appropriate reheat temperature and/or MSSM parameters.   If
the saxion abundance is comparable to the axino abundance as is
the case of the thermal regeneration, the saxion decay to ordinary
particles can cause a problem of upsetting the standard prediction
of the big bang nucleosynthesis (BBN). Such a ``saxion problem''
puts another cosmological constraints on the axino LDM models.

\section{The axino mass}

The axion supermultiplet $A=(s+ia,\tilde{a})$ consists of
the pseudo-scalar axion $a$, its scalar partner, the saxion $s$,
and its fermionic partner, the axino $\tilde{a}$.
It has the model-independent interactions with the gluon
supermultiplet $W_\alpha$
\begin{equation}
{\cal L}_A^{\rm eff} = \left.\frac{\alpha_{\rm s}}{16\pi f_a}
AW_\alpha W^\alpha \right|_F \,,
\end{equation}
where $f_a$ is the PQ symmetry breaking scale.
At present particle phenomenology, astrophysical and cosmological
observations restrict the range of $f_a$ to be
$10^9\;{\rm GeV}\lesssim f_a\lesssim 10^{12}\;{\rm GeV}$.
Then the axion mass is given by
$m_a\sim\Lambda_{\rm QCD}^2/f_a\sim 10^{-2}-10^{-5}\;{\rm eV}$.

The axino mass depends crucially on the way of supersymmetry
 breaking. In generic supergravity (SUGRA) models, it is
expected to get the typical soft mass of order $m_{3/2}\sim100$
GeV and some special arrangement, e.g. no-scale model, is needed
to allow the axino mass in the MeV scale \cite{GY,Chun:1995hc}.
Light axino can arise naturally in GMSB models where SUSY breaking
scale is lower than the PQ symmetry breaking scale
\cite{Chun:1999cq}. Let us show how the MeV axino is predicted in
GMSB models. Consider the DFSZ axion model \cite{Kim:1986ax} where
the MSSM fields are charged under the PQ symmetry. Upon the PQ
symmetry breaking, an effective K\"ahler potential between the
axion supermultiplet $A$ and the other fields $\Phi_i$ is
generated as follows;
\begin{equation} \label{Keff}
K_{eff} = e^{x_i {A+A^\dagger\over f_a} } \Phi^\dagger_i \Phi_i
\end{equation}
where $x_i$ is the PQ charge of $\Phi_i$. Taking the terms of
order $A^2$ and $\Phi_i=H_{1,2}$, one has a contribution to the
axino mass; $m_{\tilde{a}} \approx F_{H_i} v/f_a^2 \approx \mu
v^2/f_a^2 \ll$ MeV which is negligible in our context. In GMSB
models, $\Phi_i$ can be one of the hidden sector superfields, say
$\hat{X}$, which is assumed to take the vacuum expectation value;
$\langle \hat{X} \rangle = X + \theta^2 F_X$ leading to the
effective supersymmetry breaking scale, $\Lambda \equiv F_X/X
=10^4-10^5$ GeV \cite{Giudice:1998bp}. Then, one obtains the axino
and saxion mass as
\begin{eqnarray} \label{mass-as}
m_{\tilde{a}}= x_X^2  {X F_X \over f_a^2} &\approx& \left(\frac{X}{f_a}\right)^2\Lambda \,,\\
m_s^2 =2 x_X^2 {F_X^2 \over f_a^2} \,.
\end{eqnarray}
The axino mass in the range $1-10$ MeV is obtained with $X/f_a\sim
10^{-3}-10^{-4}$. These equations also give us the relation;
\begin{equation} \label{as-relation}
m_s^2 \approx 2\, m_{\tilde{a}} \Lambda
\end{equation}
leading to the saxion mass $m_s \approx 4.5-45$ GeV.

\section{The origin of cosmic axinos and cosmological constraints}

There are two known ways in which axinos are produced in the early universe.
One is the thermal production from the hot thermal bath after reheating.
The other is the non-thermal production from decays of the lightest ordinary
supersymmetric particles (LOSPs).

The decoupling temperature of axinos is estimated as \cite{Rajagopal:1990yx}
\begin{equation}
T_D \sim 10^{10}\ {\rm GeV} \left(\frac{f_a}{10^{11}\,{\rm GeV}}\right)
    \left(\frac{\alpha_s}{0.1}\right)^{-3},
\end{equation}
where $\alpha_s$ is the strong coupling constant. If the reheat
temperature $T_R$ after inflation is higher than the decoupling
temperature, the universe is overpopulated by axinos if the axino
mass is larger than a few keV. Therefore, we only consider the
case that the reheat temperature is lower than the decoupling
temperature. In this case, axinos are produced from the thermal
bath through scattering of quarks and gluons, though the number
density of them do not reach the thermal equilibrium. The amount
of axinos produced in this way, so called regeneration, is
estimated to be \cite{Covi:2001nw}
\begin{equation} \label{Omega-axino}
\Omega_{\tilde a}h^2 \approx 0.28\ \left(\frac{m_{\tilde a}}{{\rm MeV}}\right)
    \left(\frac{T_R}{10^5\,{\rm GeV}}\right)
    \left(\frac{f_a}{10^{11}\,{\rm GeV}}\right)^{-2}.
\end{equation}
Thus, for the axino with mass $1-10$ MeV to be the LDM, the
relevant range of reheating temperature is $10-100$ TeV.

The axinos from decays of LOSPs can be cosmologically interesting
when the axino mass is around the marginal value of order 10 MeV.
For this size of axino mass, the reheat temperature must be
lower than 10 TeV to suppress the thermal production
(regeneration). The amount of produced axinos is simply connected
to that of LOSPs by
\begin{equation}
\Omega_{\tilde a}h^2 = \frac{m_{\tilde a}}{m_\chi}\Omega_\chi h^2,
\end{equation}
and independent of the reheat temperature. When we take
$m_\chi=100$ GeV and $m_{\tilde a}=10$ MeV, the required value of
$\Omega_\chi h^2$ is $\sim10^4$.  Such a high  value is reached
for very large $M_{\rm SUSY}$ in the range of tens of TeV. Thus,
the non-thermal production of axinos for LDM could only be
marginally relevant.

Even though relic axinos dominantly come from regeneration, the
existence of LOSPs and their decay to axinos can produce radiative
or hadronic cascades during or after the BBN, and alter its
standard predictions on the light element abundances. To avoid this, the
mass of LOSP needs to be large enough to make its lifetime much less
than 1 sec. For example, in the case of the neutralino,
one requires $m_\chi>150$ GeV.

Let us discuss here how the accompanied saxion  can also upset the
standard prediction of the BBN, which is called ``the saxion problem''
\cite{Chang:1996ih}.
Contrary to the axino, the saxion has the axion-like couplings to the quarks,
${m_q\over f_a}s \bar{q}q$, or leptons, ${m_l\over f_a}s \bar{l}l$,
so that its life-time is much shorter than the axino LSP.
On the other hand, during the axino regeneration (\ref{Omega-axino}),
the saxions are also populated by the same amount and thus one finds
\begin{equation}
m_s Y_s \approx 10^{-9} \left(\frac{m_{s}}{{\rm MeV}}\right)
    \left(\frac{\rm MeV}{m_{\tilde{a}}}\right)
    \left(\frac{\Omega_{\tilde{a}} h^2}{0.28}\right) \, \mbox{GeV}.
\end{equation}
where $Y_s$ is the saxion number density in unit of the entropy
density. Note that this quantity is strongly constrained by the
BBN.  In the mass range $m_s \gtrsim {\cal O}(10)$ GeV, the above
equation gives  $m_s Y_s \gtrsim 10^{-5}$ GeV for $m_{\tilde{a}}
=1$ MeV.  Now that  the saxion decays mainly to bottom and charm
quarks,  one finds a strong limit on the saxion lifetime: $\tau_s
\lesssim  10^{-2}$ sec \cite{Kawasaki:2004yh}.
Specifically, the mass relation (\ref{as-relation}) gives us $m_s
\approx 14$ GeV for the axino mass $m_{\tilde{a}}\approx 1$ MeV
and $\Lambda =10^5$ GeV.   Then, the saxion lifetime,
\begin{equation}
\tau_s\approx \left[ {1\over8\pi}{m_b^2\over f_a^2}
m_s\right]^{-1} \lesssim 10^{-2}\;\mbox{sec}\,,
\end{equation}
becomes short enough to avoid the saxion problem   for
$f_a\lesssim 3\times10^{11}$ GeV. In the case of supergravity
models where one expects to get $m_s\approx 10^{2-3}$ GeV, the
saxion is free of such a problem.

\section{Axino--neutrino mixing and axino decay}

Let us now assume the generation of the bilinear superpotential
term, $H_1 H_2$, and its R-parity and lepton number violating
extension, $L_i H_2$ as a result of  the PQ symmetry breaking;
\begin{equation} \label{Weff}
W_{eff} = \mu H_1 H_2 + \epsilon_i\mu L_i H_2
\end{equation}
where $\mu$ and $\epsilon_i\mu$ carry PQ charges whose sizes are
determined by the PQ charge assignments for $H_{1,2}$ and $L_i$.
In Eq.~(\ref{Keff}), the leading terms in $A$,
\begin{equation}
K_{eff} =  {A\over f_a} [x_{H_i}  H_i^\dagger H_i + x_{L_j}
L_j^\dagger L_j ] + \cdots\,,
\end{equation}
give rise to the following axino-Higgsino and axino-neutrino mass
terms;
\begin{equation}
{\cal L}_{\rm mixing} = x_{H_1} {\mu v_2 \over f_a}\, \tilde{a}
\tilde{H}_1 + x_{H_2} {\mu v_1 \over f_a}\, \tilde{a} \tilde{H}_2
+ x_{L_i} {\epsilon_i\mu v_2 \over f_a}\, \tilde{a} {\nu}_i +
h.c.\,.
\end{equation}
For $\mu v /f_a \ll m_{\tilde{H}}$ and $\epsilon_i\mu v_2/f_a \ll
m_{\tilde{a}}$, one has the approximate mixing angles between the
axino and Higgsino or neutrino as follows;
\begin{equation}
\theta_{\tilde{a}\tilde{H}} \sim {v \over f_a} \quad \mbox{and}
\quad \theta_{\tilde{a}\nu_i} = x_{L_i} {\epsilon_i\mu v_2 \over
f_a m_{\tilde{a}}} \,.
\end{equation}

The axino-neutrino mixing derived above induces the effective
vertex of $\tilde{a}\nu_i Z$ and $\tilde{a}l_iW$ with the coupling
$\sim g\theta_{\tilde{a}\nu_i}$. This gives rise to the four-quark
operator as follows:
\begin{equation} \label{Lee}
{\cal L}_{e^+ e^-} \approx {G_F\over \sqrt{2}} \,
\theta_{\tilde{a}\nu_i} \, \bar{\nu}_i\gamma_\mu \gamma_5
\tilde{a} \, \bar{e} \gamma^\mu (2\delta_{i1} -\gamma_5)  e
\end{equation}
where we omitted the small correction due to the vector part of
the charged current.

Another important interaction to consider is the
axino-photon-neutrino vertex arising from the photino-neutrino
mixing.  The bilinear term $L_i H_2$ induces the mixing between
neutrinos and neutralinos of order $\epsilon_i$. Then the
supersymmetric anomaly coupling of axino-photon-photino leads to
the axino-photon-neutrino coupling which is written down
schematically as follows;
\begin{equation} \label{Lgamma}
{\cal L}_\gamma = {C_{a\gamma\gamma}\alpha_{\rm em}\over 8\pi f_a}
{\epsilon_i} \bar\nu_i\gamma_5\sigma_{\mu\nu}\tilde{a}F^{\mu\nu},
\end{equation}
where $C_{a\gamma\gamma}$ is an order-one parameter taking into account the
precise values of the $U(1)_{\rm em}$ anomaly and the photino-neutrino
mixing.  From the vertices (\ref{Lee}) and (\ref{Lgamma}), we get
the following decay widths of the axino;
\begin{eqnarray}
\Gamma_{\nu_i e^+ e^-} &=& ~{G_F^2 m^5_{\tilde{a}}\over 192 \pi^3}
\,\theta_{\tilde{a}\nu_i}^2 [{1\over4} +\delta_{i1}]  \nonumber \\
\Gamma_{\nu_i\gamma} &=& ~ {C_{a\gamma\gamma}^2\alpha_{\rm em}^2 \over
(16\pi)^3} { m^3_{\tilde{a}} \over f_a^2 } \,\epsilon_i^2\,.
\end{eqnarray}
Let us first note that the photon mode is suppressed by
$\alpha_{\rm em}^2$ compared to the $e^+e^-$ mode;
\begin{equation}
{\Gamma_{\nu\gamma} \over \Gamma_{\nu e^+ e^-} } \approx
{3 C_{a\gamma\gamma}^2\alpha_{\rm em}^2 \over 32 G_F^2 \mu^2 v^2} \approx 10^{-4}
\end{equation}
for $\mu/C_{a\gamma\gamma} = 100$ GeV.
It is smaller than the internal bremsstrahlung process of $e^+e^-$ mode
which is suppressed by $\alpha_{\rm em}$ and also produces a direct $\gamma$-ray.
This is enough to be consistent with the observations of
the MeV $\gamma$-ray spectrum \cite{Beacom:2004pe}.
Then, the axino decay is determined by the process
$\tilde{a}\to\nu e^+e^-$ whose lifetimes is given by
\begin{eqnarray}
\tau_{\tilde{a}} &\approx& 10^{26}\sec \left(\frac{1\mbox{
MeV}}{m_{\tilde{a}}}\right)^3
\left(\frac{f_a}{10^{11}\mbox{ GeV}}\right)^2
\left(\frac{10^{-7}}{|x_{L}\epsilon|}\right)^2
\left(\frac{100\mbox{ GeV}}{\mu}\right)^2 \label{tau-axino}
\end{eqnarray}
which is in the right range to explain the observation
(\ref{primo}) consistently with the neutrino data as will be shown
in the following section.

\section{Consistency with the  neutrino data and experimental signatures}

One of the interesting aspect of R-parity violation is that it can
be the origin of the observed  neutrino masses and mixing
\cite{LeeKang}.  The general superpotential allowing R-parity and
lepton number violation includes the following bilinear and
trilinear  terms;
\begin{equation} \label{WRpV}
 W_{Rp}= \epsilon_i\mu L_i H_2 + {1\over2}\lambda_{ijk} L_i L_j E^c_k
 +\lambda'_{ijk}
 L_i Q_j D^c_k \,.
 \end{equation}
According to the observation of Ref.~\cite{Hooper:2004qf}, an
appropriate life time of the axino decay $\tilde{a}\to \nu_{\mu,
\tau} e^+ e^-$ can arise with trilinear R-parity violating
couplings $\lambda_{211, 311} \sim 0.1$.
Such trilinear couplings can generate the 2-3 components of the
neutrino mass matrix;
\begin{equation}
 M^\nu_{ij} \approx {1\over 8\pi^2} \lambda_{i11}\lambda_{j11}
{m_{e}^2 \mu \tan\beta \over m_{\tilde{e}}^2 }
\end{equation}
where $\tan\beta \equiv \langle H_2^0 \rangle / \langle H_1^0
\rangle$ and $m_{\tilde{e}}$ is the selectron  soft mass. While
the charged-current and  $e$--$\mu$--$\tau$ universality put the
bound $\lambda_{i11} \lesssim 0.1 (m_{\tilde{e}}/200 \mbox{ GeV})$
\cite{Barger:1989rk}, the above one-loop mass can reach the
observed atmospheric neutrino mass scale $m_\nu \approx 0.05$ eV
only for an extreme value of $\mu\tan\beta \approx 50$ TeV taking
the boundary value of $\lambda_{i11}=0.1\,
(m_{\tilde{e}}/200\,{\rm GeV})$. In order to generate the other
components of the neutrino mass matrix,  one needs to introduce
some other trilinear couplings such as $\lambda_{i22, j33}$ which
induce $M^\nu_{11}, M^\nu_{12}$ and $M^\nu_{13}$ through the
combinations of $\lambda_{1jj}\lambda_{1jj}$,
$\lambda_{133}\lambda_{233}$ and $\lambda_{122}\lambda_{322}$,
respectively.   Then, appropriate neutrino masses can be obtained
for the  trilinear couplings, $\lambda_{i22}\sim 10^{-4}$ and
$\lambda_{i33}\sim 10^{-5}$, where the small ratios
$\lambda_{i22}/\lambda_{i11}$ and $\lambda_{i33}/\lambda_{i11}$
are dictated by the factors of $m_e/m_\mu$ and $m_e/m_\tau$,
respectively. Such a hierarchy among $\lambda_{ijj}$ appears {\it
ad-hoc} considering the usual hierarchy in the quark and lepton
Yukawa couplings.

Nevertheless, if there exits the trilinear coupling
$\lambda_{i11}$ of order 0.1, they leads to a remarkable
experimental signature of resonant single sneutrino production in
the future linear collider
\cite{Dimopoulos:1988jw,Abdallah:2002wt}, non-observation of which
would rule out the axino LDM decaying through the trilinear
couplings.

The observed neutrino masses and mixing can be more naturally
explained if one invokes the presence of the bilinear term of the
order $10^{-6}$  \cite{LeeKang}.
The bilinear R-parity violation generates neutrino masses at
tree-level through the neutrino--neutralino mixing.  In addition
to the $\epsilon_i$ term in the superpotential (\ref{WRpV}), the
scalar potential also contains the R-parity violating bilinear
soft terms as follows;
\begin{equation} \label{VRpV}
 V_0 = m^2_{L_i H_1} L_i H_1^\dagger +  B_i L_i H_2  + h.c.,
\end{equation}
where $B_i$ is the dimension-two soft parameter.  Generically, one
has $B_i=\epsilon_i \tilde{B} \mu$ with a dimension-one soft
parameter $\tilde{B}$ for the $\mu$ term, and the soft
mass-squared $m^2_{L_i H_1}$ contains the supersymmetric term
$\epsilon_i \mu^2$. Upon the electroweak symmetry breaking, the
sneutrino field gets nontrivial vacuum expectation value;
\begin{equation}
  \frac{\langle \tilde{\nu_i} \rangle}
 {v_1 } = - {m^2_{L_i H_1} + B_i \tan\beta
          \over m^2_{\tilde{\nu}_i} } \,,
\end{equation}
which is expected to be of order $\epsilon_i$ up to the soft mass
dependence. These bilinear parameters induce mixing between
neutrinos and neutralinos. For the small mixing mass, the week-scale
seesaw with heavy neutralino mass scale $\sim 100$ GeV leads to the
well-known neutrino mass matrix at tree-level;
\begin{equation} \label{Mtree}
M^\nu_{ij} = - {M_Z^2 \over F_N}  \xi_i \xi_j \cos^2\beta
\end{equation}
where $\xi_i \equiv \epsilon_i - \langle \tilde{\nu_i}
\rangle/v_1$ and $F_N= M_1M_2/M_{\tilde{\gamma}}+ M_Z^2
\cos{2\beta}/\mu$ with $M_{\tilde{\gamma}} = c_W^2 M_1 + s_W^2
M_2$.  From Eq.~(\ref{Mtree}), one obtains the size of
$|\xi|=\sqrt{\sum_i|\xi_i|^2}$ consistently with the atmospheric
neutrino mass scale as follows;
\begin{equation}
|\xi| = 0.7\times10^{-6} \frac{1}{\cos\beta}
\left(\frac{F_N}{M_Z}\right)^{1/2} \left(\frac{m_{\nu}}{0.05\,{\rm
eV}}\right)^{1/2} \,.
\end{equation}
This  is compatible with the axino lifetime relation
(\ref{tau-axino}) for $\xi_i \sim \epsilon_i$.  Note that the
smaller neutrino mass explaining the solar neutrino oscillation
can come from one-loop diagrams involving the trilinear couplings
of order, $\lambda_{i33}, \lambda'_{i33} \sim 10^{-4:-5}$.

Let us finally remark that the bilinear R-parity violation leads
to a distinct prediction on the lepton number violating decays of
the lightest neutralino $\chi$ in the future colliders. The mass
matrix of the form (\ref{Mtree}) enables us to determine the
relation   $5 |\xi_1| \lesssim |\xi_2| =|\xi_3|$ from  the
neutrino data on the mixing angles. As the parameters $\xi_i$
determine also the couplings of the R-parity violating processes;
$\chi \to l_i^{\pm} W^{\mp}$,  the  above mixing angle relation
can be tested  in the decay of the neutralino whose branching
ratios satisfies ${\rm Br}(eW): {\rm Br}(\mu W) : {\rm Br}(\tau W)
= |\xi_1|^2: |\xi_2|^2: |\xi_3|^2$ \cite{LeeKang}. It is
intriguing to note that future colliders can provide an indirect
test for either scenario of the axino LDM decaying through
$\lambda_{i11}$ or $\epsilon_i$.

\section{Conclusion}

The axino with the mass in the $1-10$ MeV range is a good candidate
for the LDM, which not only constitutes CDM but also explains the
observed 511 keV $\gamma$-rays from the galactic bulge through its
decay. The desired mass of the axino can be realized in certain
supergravity models with some special arrangements, e.g., no-scale
K\"ahler potential, or in gauge-mediated SUSY breaking models. The
origin of relic axinos can be either the thermal production from the
thermal bath after reheating or the non-thermal production from the
LOSP decays. Both require a rather low reheat temperature
$T_R\sim10-100$ TeV.

The long lifetime of the axino is a result of the R-parity
violation and the suppression of axino interactions with ordinary
particles by the PQ scale. As is well-known, the small violation
of R-parity by bilinear terms is an attractive option for
generating the neutrino masses and mixing. We found an interesting
fact that the same small R-parity violating bilinear terms can
explain the observed 511 keV $\gamma$-rays as well as the observed
neutrino mass matrix consistently within the current observational
bounds and the reasonable choice of model parameters. This
connection has a virtue that the explanation of neutrino masses
and mixing by R-parity violating bilinear terms has testable
predictions in the future colliders, thereby provides an indirect
test of decaying axino LDM. The LDM is an very attractive idea in
that if it turns out to be true, the morphology of 511 keV
gamma-rays will serve as a good probe of the dark matter halo
density profile. The decaying LDM models require more curspy
density profile to fit the observed morphology of 511 keV
$\gamma$-rays from the galactic bulge than the annihilation
models. We expect this leads to interesting astrophysical
implications \cite{Hooper:2003sh,Ascasibar:2005rw}.

\begin{acknowledgments}
This work is supported by the Science Research Center Program of
the Korean Science and Engineering Foundation (KOSEF) through
the Center for Quantum Spacetime (CQUeST) of Sogang University
with grant No.~R11-2005-021 and
the grant No.~R01-2004-000-10520-0 from the Basic Research Program
of the KOSEF (H.B.K.).
\end{acknowledgments}

\end{document}